\documentclass[runningheads]{llncs}
\usepackage{graphicx}
\usepackage{amsmath}

\newcommand{\LL}{\ensuremath{\mathbf{L}}}

\begin{document}

\renewcommand{\arraystretch}{1.4}
\setlength{\tabcolsep}{0.3em}

\title{Redundancy Reduction in Semantic Segmentation of 3D Brain Tumor MRIs}
\titlerunning{Redundancy Reduction in Semantic Segmentation of 3D Brain Tumor MRIs}
\author{Md Mahfuzur Rahman Siddiquee, Andriy Myronenko}
\authorrunning{M. Siddiquee}
\institute{NVIDIA, Santa Clara, CA \\ \email{mrahmans@asu.edu, amyronenko@nvidia.com} }
\maketitle              
\begin{abstract}
Another year of the multimodal brain tumor segmentation challenge (BraTS) 2021 provides an even larger dataset to facilitate collaboration and research of brain tumor segmentation methods, which are necessary for disease analysis and treatment planning. A large dataset size of BraTS 2021 and the advent of modern GPUs provide a better opportunity for deep-learning based approaches to learn tumor representation from the data. In this work, we maintained an encoder-decoder based segmentation network, but focused on a modification of network training process that minimizes redundancy under perturbations.  Given a set trained networks, we further introduce a confidence based ensembling techniques to further improve the performance. We evaluated the method on BraTS 2021 validation board,  and achieved  0.8600, 0.8868 and 0.9265 average dice for enhanced tumor core,  tumor core and whole tumor, respectively. Our team (NVAUTO) submission was the top performing in terms of ET and TC scores and within top 10 performing teams in terms of WT scores.

\end{abstract}

\section{Introduction}

This year, 2021 multimodal brain tumor segmentation challenge (BraTS)~\cite{brats21}, boasts a large dataset of 2000 cases.  Long gone the days of a modest couple of hundred image datasets~\cite{BratsAll2018,Myronenko18}. With a large data size comes more confidence in the achieved solution to be better generalizible in clinical practice. The dataset includes data from the previous year challenges, which also implies that some cases might be very dated (of low image quality), but this can be considered a positive feature to ensure the robustness of the solution.  A large data size, in theory, should lead to a more accurate solution, but also requires a large computational resource.  Thankfully, modern GPUs, such as as Nvidia A100 (RTX generation), have substantially improved over the last years in terms of speed, performance and GPU memory sizes. It is also interesting to note of the paradigm change, that it is no longer necessary to motivate the necessity of deep learning based approach here, as every  solution in this challenge is probably deep learning based. 

Gliomas are primary brain tumors, which originate from brain cells, whereas  secondary tumors metastasize into the brain from other organs. 
Gliomas can be of low-grade  (LGG) and high-grade (HGG) subtypes. High grade gliomas are an aggressive  type of malignant brain tumor that grow rapidly, usually require surgery and radiotherapy and have poor survival prognosis. Magnetic Resonance Imaging (MRI) is a key diagnostic tool for brain tumor analysis, monitoring and surgery planning. Usually, several complimentary 3D MRI modalities are acquired - such as T1, T1 with contrast agent (T1c), T2 and Fluid Attenuation Inversion Recover (FLAIR) - to emphasize different tissue properties and areas of tumor spread.  For example the contrast agent, usually gadolinium, emphasizes hyperactive tumor subregions in T1c MRI modality.  

Multimodal Brain Tumor Segmentation Challenge (BraTS) aims to evaluate state-of-the-art methods for the segmentation of brain tumors by providing a 3D MRI dataset with ground truth tumor segmentation labels annotated by physicians~\cite{BratsAll2018,brats1,brats2,brats3,brats4}. This year, BraTS 2021 training dataset included 1251 cases, each with four 3D MRI modalities (T1, T1c, T2 and FLAIR) rigidly aligned, resampled to 1x1x1 mm isotropic resolution and skull-stripped. The input image size is 240x240x155. The data were collected from multiple institutions, using various MRI scanners. Annotations include 3 tumor subregions: the enhancing tumor, the peritumoral edema, and the necrotic and non-enhancing tumor core.  The annotations were combined into 3 nested subregions: whole tumor (WT), tumor core (TC) and enhancing tumor (ET), as shown in Figure~\ref{fig:seg}. Two additional datasets without the ground truth labels were provided for validation and testing.  The validation dataset (219 cases) allowed multiple submissions and was designed for intermediate evaluations. The testing dataset (530 cases) will be analysed blindly using the docker submission, and is used to calculate the final challenge ranking.  

In this work, we describe our approach for  3D brain tumor segmentation from multimodal 3D MRIs and participate in BraTS 2021 challenge.

\section{Related work}
\label{sec:relatedwork}

Previous years of BraTS challenge winners include our work in 2018~\cite{Myronenko18}, using encoder-decoder based network with asymmetrically large encoder.  In 2019, Jiang et al.~\cite{jiang2019brats} used a similar encoder-decoder architecture, but added a second network to refine the segmentation in a cascade fashion. In 2020,  Isensee et al.~\cite{isensee2020nnunet} proposed  modifications to their nnUnet work. The authors continue to demonstrate that a standard encoder-decoder based architecture can achieve state of the art results, given the best practices of all components of the segmentation workflow (pre-processing, normalization, augmentations, ensembling).  Here, we generally follow our encoder-decoder based network from~\cite{Myronenko18} as a backbone, but modify the training and ensembling strategies.  

\section{Methods}

\label{sec:methods}
We used SegResNet from MONAI~\cite{monai,Myronenko18} with encoder-decoder based CNN architecture. We switched to  Instance Normalization~\cite{Instance16}, as it has a lightly lower memory footprint compared to Group Normalization (GN)~\cite{Wu18}, for equivalent performance. Briefly speaking, the encoder part uses  ResNet~\cite{He16}  blocks, where each block consists of two convolutions with normalization and ReLU, followed by additive identity skip connection. We  progressively downsize image dimensions by 2 and simultaneously increase feature size by 2.  Each decoder level begins with upsizing followed by an addition of encoder output of the equivalent spatial level. The end of the decoder has the same spatial size as the original image, and the number of features equal to the initial input feature size, followed by 1x1x1 convolution into 3 channels and a sigmoid function. See Table~\ref{tab:encoderdecoder} for the details.

\begin{table}
	\centering
	\caption{Backbone network structure, where IN stands for group normalization, Conv - 3x3x3 convolution, AddId - addition of identity/skip connection.}
	\label{tab:encoderdecoder}
	\begin{tabular}{|l|c|c|}
		 \hline
		Name & Ops & Repeat    \\ \hline
		InitConv & Conv & 1     \\
		EncoderBlock0 & IN,ReLU,Conv,IN,ReLU,Conv, AddId & 1    \\
		EncoderDown1 & Conv stride 2  &  1  \\
		EncoderBlock1 & IN,ReLU,Conv,IN,ReLU,Conv, AddId& 2     \\
		EncoderDown2 & Conv stride 2 &  1  \\
		EncoderBlock2 & IN,ReLU,Conv,IN,ReLU,Conv, AddId& 2     \\
		EncoderDown3 & Conv stride 2 & 1     \\
		EncoderBlock3 & IN,ReLU,Conv,IN,ReLU,Conv, AddId& 4   \\
		DecoderUp2 & UpConv, +EncoderBlock2 &  1     \\
		DecoderBlock2 & IN,ReLU,Conv,IN,ReLU,Conv, AddId & 1     \\
		DecoderUp1 & UpConv, +EncoderBlock1 &  1     \\
		DecoderBlock1 & IN,ReLU,Conv,IN,ReLU,Conv, AddId & 1    \\
		DecoderUp0 & UpConv, +EncoderBlock0 &  1    \\
		DecoderBlock0 & IN,ReLU,Conv,IN,ReLU,Conv, AddId & 1    \\
		DecoderEnd & IN,ReLU, Conv1, Sigmoid &  1     \\
		\hline
		
	\end{tabular}
\end{table}

 \subsection{Redundancy Reduction}

We would like to encourage structure on the learned feature representations of our network. Firstly, we want to enforce invariance of the result under perturbations.  Generally, it is common to use image augmentations, such as intensity modifications or spatial transforms, to implicitly enforce network robustness to such image perturbations. However, this alone does not often lead to a similar feature representation under perturbations, even if the final segmentation masks are similar. In semi-supervised learning, people adopted such method as contrasting coding of SimCLR~\cite{chen2020simple} to enforce consistency of learned features under perturbations. Contrastive coding  requires large number of negative samples/batches, which is too computationally expensive for 3D images. Secondly, we  also want another useful property of the learned image features: minimal redundancy. Loosely speaking, we want different samples to have different learned representation with minimal information overlap.  For these reason, we adopted the loss term from Barlow Twins~\cite{zbontar2021barlow}, which is:
  \begin{equation}
 \label{eq:loss}
 \LL_{BT} =\LL_{invariance} + \LL_{redundancy}
 \end{equation}

Unlike the original paper concerning classification, our adaptation is for the segmentation which requires some modifications.   For example, when training with a 3D batch size of 1, we copy it and augment it each copy independently, forming a batch of 2. 
For augmentations we use, both intensity and spatial augmentations.  We then attach a projection branch to the output feature dimension (one level before the final normalization). The projection branch first average pools the features by a factor of 16, for two reasons: the output features are of the original image size, and it's too computationally expensive to use them directly, secondly spatial augmentation transforms would produce geometrical image misalignment between the two images, and average pooling ensures that the pooled features are roughly of the same spatial content. After pooling the newly formed features are reshaped to form a large batch of 1D features. 

At the stage we proceed with the projection branch (3 layer MLP) and compute the empirical cross-correlation matrix $C$ between all combinations of spatial regions~\cite{zbontar2021barlow}:
  \begin{equation}
 \label{eq:cmat}
 C_{ij} =\frac{\sum_b z^A_{b,i} z^B_{b,j}}{\sqrt{\sum_b (z^A_{b,i})^2} \sqrt{\sum_b (z^B_{b,j})^2}}   
 \end{equation}
where b indexes batch samples and i,j index the vector dimension of the networks’ outputs. C is a square matrix with
size the dimensionality of the network’s output, and with
values comprised between -1 (i.e. perfect anti-correlation)
and 1 (i.e. perfect correlation).

Unlike the original work, which uses a batch of many different 2D images, our cross-correlation matrix is based on the various regions of the same  image under permutations. Based on the cross-correlation matrix,  we can compute a differentiable loss:
  \begin{equation}
 \label{eq:loss2}
 \LL_{BT} =\sum (1-C_{ii})^2 + \lambda \sum_{i\neq j}C_{ij}^2   
 \end{equation}

where the first part enforces invariance of learned feature representations from the same spatial regions under perturbations, and second term reduces redundancy between different spatial regions. We used $\lambda=0.005$ weight, to balance the parts, as proposed in the original paper. The projection branch is discarded during inference. 
We use this loss in additional to the main soft Dice loss~\cite{Milletari16}:  
   \begin{equation}
  \label{eq:dice}
  \LL_{dice}  = 1 - \frac{2*\sum p_{true} * p_{pred} }{\sum p_{true}^2 + \sum p_{pred}^2 + \epsilon}   
  \end{equation}
where $p_{true}$ is a  ground truth binary mask, and $p_{pred}$ is the predicted probability (after sigmoid) per class. 

 \subsection{Confidence based ensembling}
Given a trained set of models, ensembling is a popular technique to boost the performance, typically by averaging the probabilities of predictions. Alternatively one can employ a geometrical mean (instead of standard mean) to combine the probabilities. We take it one step further and design a technique to adaptively average the probability maps. 

Assuming we know the confidence of the solution/prediction result of each model for a given image,  then we could use the subset of the probability maps with the top confidence for ensembling (or we can use the weight mean of all such maps). We found that using the top N/2 probabilities (with equal weights) generally gives the best performance, where N is the number of all maps before the ensembling. 

Estimation of the confidence measure of the given segmentation probability map is non-trivial. We propose a heuristic for such estimate as the average probability value of the segmented region. Such heuristic is based on the observation that given two probability maps of the same final segmentation volume, the one with the highest average probability is usually more accurate. This assumption is not true when different models produce substantially different region sizes, or if the trained model were very different to compare directly. In our experiment, we found such heuristic consistently improves the ensembling performance.

 \subsection{Optimization}
 We use AdamW optimizer with initial learning rate of $ \alpha_{0} = 10^{-4}$ and progressively decrease it according to Cosine schedule. We use  $10^{-5}$ weight decay and no dropout.  

  \subsection{Data prepossessing and augmentation}
  We normalize all input images to have zero mean and unit std (based on non-zero voxels only).  We also apply a random axis mirror flip (for all 3 axes) with a probability $0.5$. After this we copy the image, and apply the intensity and spatial transforms.  

 \section{Results}
 \label{sec:results}
 
 \begin{figure}[t] 
 	\centering
 	\includegraphics[clip=true, trim=0pt 0pt 0pt 0pt, width=0.99\textwidth]{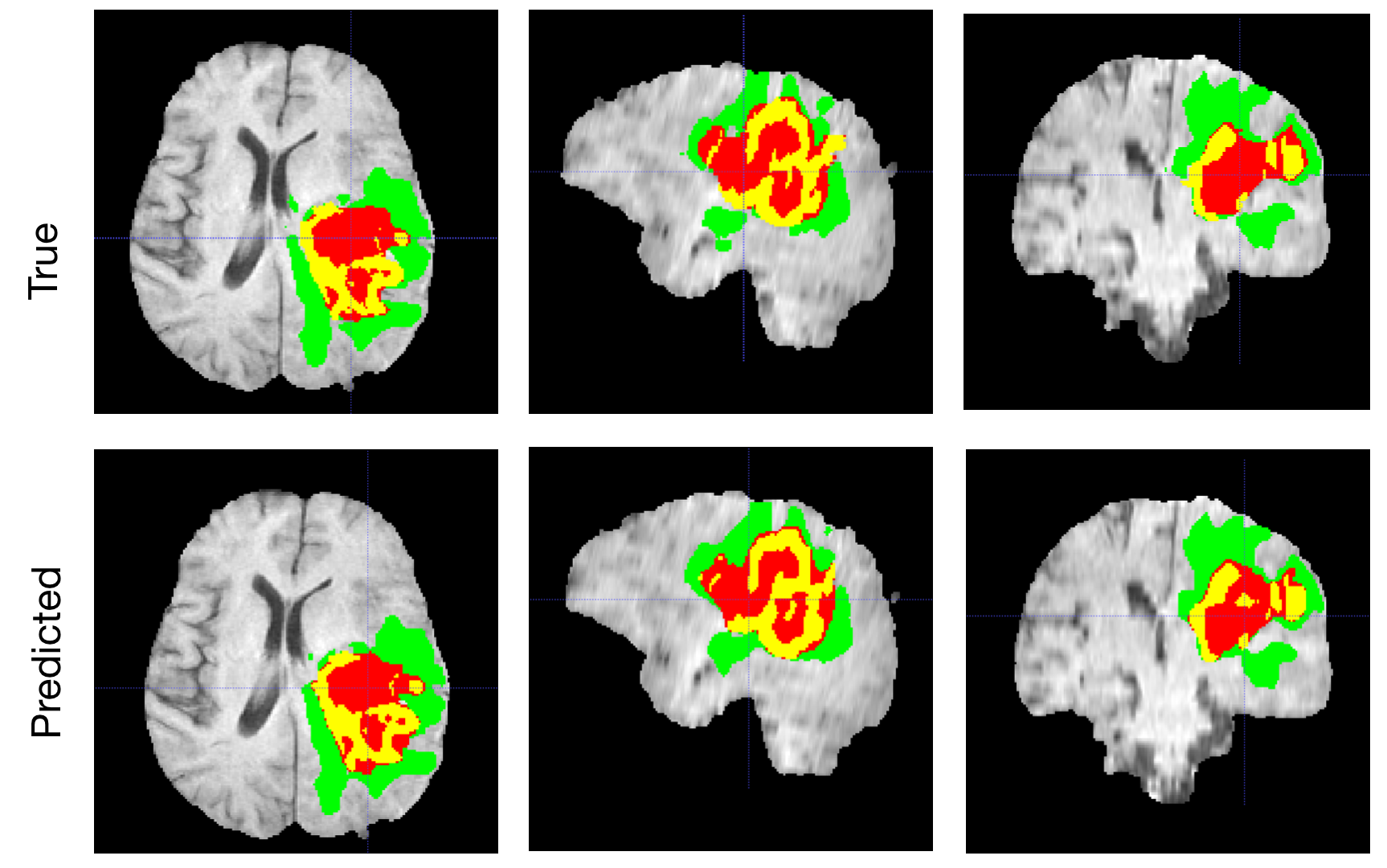}
 	\caption{A typical segmentation example with true and predicted labels overlaid over T1c MRI axial, sagittal and coronal slices.  The whole tumor (WT) class includes all visible labels (a union of green, yellow and red labels), the tumor core (TC) class is a union of red and yellow, and the enhancing tumor core (ET) class is shown in yellow (a hyperactive tumor part). }
 	\label{fig:seg}
 \end{figure}

 We implemented our network in MONAI~\footnote{https://github.com/Project-MONAI/MONAI} and trained it on 4 Nvidia V100 32GB GPUs using 5-fold cross-validation. During training we used a random crop of size 192x192x144, which ensures that most image content remains within the crop area. We concatenated  4 available 3D MRI modalities into the 4 channel image as an input. We train it on 4 GPU for 300 epochs in about 24hr. The output of the network is 3 nested tumor subregions (after the sigmoid). 
 
\begin{figure}[ht!]
     \includegraphics[clip=true, trim=0pt 0pt 0pt 0pt, width=0.32\textwidth]{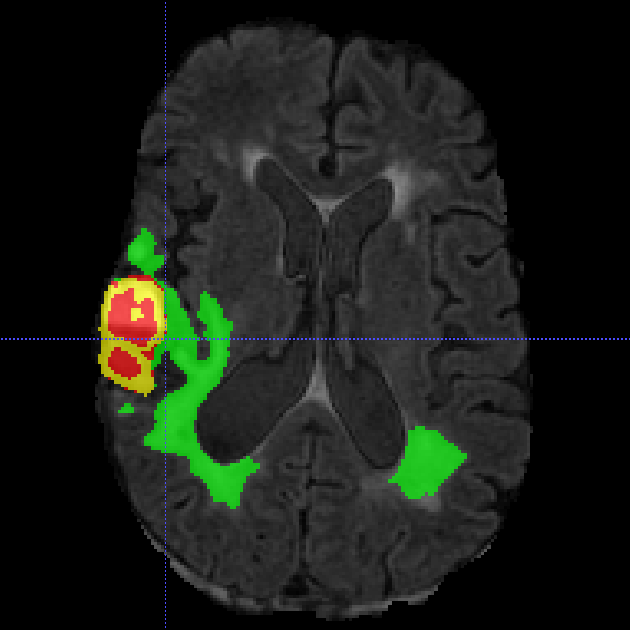}\hfill
     \includegraphics[clip=true, trim=0pt 0pt 0pt 0pt, width=0.32\textwidth]{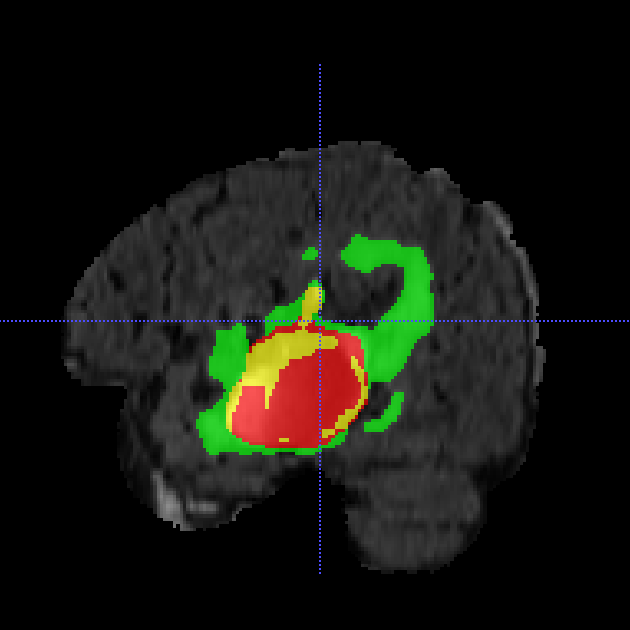}\hfill
     \includegraphics[clip=true, trim=0pt 0pt 0pt 0pt, width=0.32\textwidth]{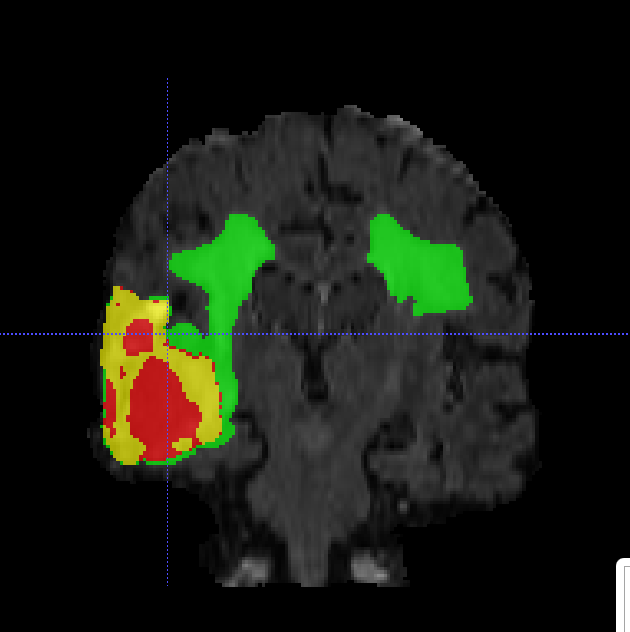}\hfill
    \caption{An example of the incorrectly over-segmented result. The whole tumor (WT) in green is incorrectly spilled over on the right side of the brain, most likely because the underlying  MRI (Flair modality) has high intensity regions which usually corresponds to the tumor.  }
     \label{fig:segincorrect}
\end{figure}

 Typically, the output result, even of a single model, is accurate as shown in Figure~\ref{fig:seg}, however several cases still remain  segmented imprecisely. Figure~\ref{fig:segincorrect} shows an example of an incorrectly over-segmented whole tumor (WT) region, which is spilled over on the right side of the brain, most likely because the underlying MRI (Flair) has substantially higher intensity values in that region.  More variability in training examples might have helped to solve the issue, or integration of anatomical knowledge of e.g. "symmetrical highlights around ventricles are unlikely to be a tumor" , but  such information  is rather complicated to put inside  of the network.

We report the cross-validation results using our own 5-fold splits of the training portion of the data in Table~\ref{tab:validfolds} and the evaluated results of the BraTS 2021 validation set (219 cases) in  Table~\ref{tab:valid}. To get the validation results, we uploaded our segmentation masks to the BraTS 2021 server for evaluation, since the ground truth is not provided.  Compared to the validation results, our own cross-validation per fold results are very similar, but 1-2\% higher. A slight disagreement is normal, and due to a variability of data in the validation sets, and due to the fact that we are able to select the best checkpoint using our own 5-folds.

\begin{table}
	\centering
	\caption{5-fold cross-validation results using our own random data splits on BraTS 2021 training dataset. Mean Dice: EN - enhancing tumor core, WT - whole tumor, TC - tumor core.}
	\label{tab:validfolds}
	\begin{tabular}{l|c|c|c}
		\hline

		Fold  & ET & TC & WT  \\ \hline
		0 & $0.8759\pm0.007$ & $0.9152\pm0.009$ & $0.9332\pm0.011$  \\
		1 & $0.8993\pm0.012$ & $0.9213\pm0.007$ & $0.9379\pm0.005$  \\
		2 & $0.8880\pm0.006$ & $0.9220\pm0.013$ & $0.9313\pm0.014$  \\
		3 & $0.8902\pm0.008$ & $0.9258\pm0.007$ & $0.9351\pm0.008$  \\
		4 & $0.8847\pm0.011$ & $0.9198\pm0.013$ & $0.9374\pm0.005$  \\

		\hline
	\end{tabular}
\end{table}

\begin{table}
	\centering
	\caption{BraTS 2021 validation dataset results. Mean Dice and Hausdorff95 measurements of the proposed segmentation method. EN - enhancing tumor core, WT - whole tumor, TC - tumor core.}
	\label{tab:valid}
	\begin{tabular}{l|c|c|c|c|c|c}
		\hline
		& \multicolumn{3}{c|}{Dice} & \multicolumn{3}{c}{Hausdorff95 }  \\ \hline
		Validation dataset & ET & TC & WT & ET & TC & WT \\ \hline
		Our result (NVAUTO) & 0.8600 & 0.8868 & 0.9265 & 9.0541 & 5.8409 & 3.6009 \\
		\hline
	\end{tabular}
\end{table}

\section{Discussion and Conclusion}
 \label{sec:conclusion}
In this work, we described a semantic segmentation network for brain tumor segmentation from multimodal 3D MRIs for BraTS 2021 challenge. 
We have introduced  redundancy reduction to enforce structure on the learned features, and demonstrated confidence based ensembling technique.  
Our BraTS 2021 final validation dataset results were 0.8600, 0.8868 and 0.9265 average dice for enhanced tumor core,  tumor core and whole tumor, respectively. As of Aug 21, 2021, when the validation server of Brats 2021 stopped accepting new submissions, our results represented the top performing team in terms of the ET and TC dice scores, and were within the top 10 teams in terms of the WT dice score.

\bibliographystyle{splncs04}
\bibliography{paper}

\end{document}